# A SYSTEMATIC REVIEW OF RESEARCH ON THE USE AND IMPACT OF TECHNOLOGY FOR LEARNING CHINESE

Angelina MAKSIMOVA
Peking University, Graduate School of Education, China
angelina.maksimova@gmail.com

## ABSTRACT

*In light of technological development enforced by the Covid-19 pandemic, learning Chinese has become more digitalised. Confucius institutes went online and now follow 2021 to 2025 Action Plans for the Construction of Teaching Resources for International Chinese Education and International Chinese Online Education. New ways of learning Chinese emerged, such as educational games and intelligent tutoring systems (ITS), some of them based on artificial intelligence. The aim of this systematic review is to examine recent (from 2017 to 2022) research published in ScienceDirect and Scopus databases on the use and impact of educational games and ITS in Chinese language learning. A total of 29 selected studies were analysed. Based on the overall findings, games and ITS are effective tools for Chinese learning, that impact students' motivation, self-efficacy progress, and learning satisfaction. However, more in-depth research should explore how games and ITS can best be implemented to teach Chinese to foreigners.*

## KEYWORDS

*Chinese language learning, games, educational games, intelligent tutoring system, emotional intelligent tutoring system, flipped classroom, artificial intelligence.*

## 1. INTRODUCTION

Teaching language to foreigners is a good soft power tool, which is why a lot of countries invest in culture and education. In 2019, China was ranked 17th by the Education sub-index of the Soft Power 30 index, following western economies such as the US, the UK, Germany, Sweden, Denmark, Belgium, Australia, France, Netherlands, Canada, Switzerland, Italy, Norway, Finland, as well as South Korea and Japan. Western countries exercise soft power through an extensive network of cultural centres: Alliance Francaise, British Council, Deutscher Akademischer Austausch Dienst (DAAD), The Cervantes Institute (Maksimova, 2021). China's economic growth, a rich culture that includes the highest number of UNESCO World heritage sites and success in the Olympics contribute to the popularisation of the Chinese language.

China exercises cultural (文化软实力) and educational (教育软实力) soft power and has the objective of training foreign talents to "know China" （之花), "be friendly towards China" (优化), and "love China" (爱花). Before the Covid-19 pandemic China had become an attractive destination for studies and research. A Chinese language proficiency test (汉语水平考试; HSK) was established in 1990 as the threshold for enrolling international students in individual programmes. From 2004 to 2020, Confucius Institutes (孔子学院) opened 548 Confucius Institutes and 1,193 centres in schools with 46,700 full-time and part-time teachers in 154 countries around the world (Confucius Institute Headquarters, 2020). Learning Chinese has become more and more digitalised. From 2021 to 2025, the Action Plan for the Construction of Teaching Resources for International Chinese Education and the Action Plan for International Chinese Online Education is being implemented. In addition, games and intelligent tutoring systems (ITS) have been developed for learning Chinese.

Gamification techniques have been adopted in language learning by making the process more active and participative. Computer games, not only educational ones, have also proven to expand a learner's vocabulary. Research on gamification techniques in mainly language learning focuses on proving the efficiency of developed games on learners. In most cases, games or ITS developed by authors are aimed at students learning Chinese and are effective according to pre-test and post-test results. However, the field is still missing a comprehensive review that takes games, gamification techniques, and ITS into consideration. This review thus aims to review which games, gamification techniques, or ITS can be used in learning Chinese and what their effects on students are, based on reviewing the previous publications from databases like ScienceDirect and Scopus. The review aims to complement the previous reviews (Hung et al., 2018; Lai & Bower, 2019) on learning languages by focusing specifically on Chinese language learning using not only games, gamification, but also ITS. At present, there is no relatively comprehensive systematic review on these three methods in learning Chinese. Therefore, this review fills this gap.

The research questions are as follows:

(1) Which technologies (games, gamification, ITS) are used so far in learning Chinese?

(2) What characteristics (motivation, self-efficacy, progress, effectiveness, learning satisfaction) these technologies are enhancing?

(3) What is the future research in this field?

## 2. METHODOLOGY

The methodology of this paper represents a systematic review to identify the research papers based on the keywords in two databases - ScienceDirect and Scopus. The review was performed during the period from 2016 to May 2022 using the following keywords: Chinese language AND game, Chinese language AND gamification, Chinese language AND intelligent tutoring system. Most of the papers were found in ScienceDirect – 6 857 studies (Table 1). In the Scopus, 181 papers were detected. Thus, altogether 7 038 publications were detected in two databases. Since most of the papers focused on the use of games and ITS in learning in general, the titles, abstracts, and introductions were reviewed, and only 29 papers focusing on Chinese language learning were included into the in-depth analysis. Therefore, the paper was included if it matched the corresponding period, i.e., from 2016 up to May 2022; if the intervention involved the use of a game or gamification, or ITS; if it focused on the learning of Chinese language; if the study was written in English; and if it could be accessed through ScienceDirect and Scopus database. If the study was not accessible with the institutional login of Peking University, it was not included in the review (six in total). If the study was found in both databases, it was counted only once as a ScienceDirect study.

Table 1. Numbers of studies by the database and searched words.

| | ScienceDirect | | Scopus | |
|---|---|---|---|---|
| | All results | Studies selected for deeper analysis | All results | Studies selected for deeper analysis |
| Chinese language AND game | 6377 | 4 | 155 | 15 |
| Chinese language AND gamification | 272 | 5 | 18 | 2 |
| Chinese language AND intelligent tutoring system | 208 | 3 | 8 | 0 |
| | **6857** | **12** | **181** | **17** |

As is seen from the Annex 1, the majority of the studies (26 out of 29 studies selected for deeper analysis) focus on students' motivation (five studies measure motivation specifically, but with different techniques). Seventeen studies address effectiveness, thirteen - self-efficacy, ten - students' progress, eight - learning satisfaction. Xu et al. (2021) address all five issues. The analysis of the included studies is consistent with the previous research on games and gamification, and ITS in language learning: (1) games, gamification, and ITS have been used in language learning, also in learning Chinese (2) the majority of studies address students' motivation; (3) most studies are about effectiveness, self-efficacy, students 'progress, learning satisfaction. Based on the overall findings, games and ITS are effective and motivating tools for language learning, also Chinese, but more in-depth research should explore how they can best be implemented for Chinese language teaching to foreigners.

## 3. FINDINGS

### 3.1. Gaming

In language learning, games help to dive into the language learning environment or use the foreign language for interaction (Chen et al., 2020). Gamification in Massive Open Online Courses (MOOCs) has supported Chinese language learning for a long time (Metwally&Yining, 2017). The majority of studies, especially those designed for interventions into the learning process in the classroom setting, are aimed at proving gamification effectiveness through games' impact on students' motivation and attitude towards learning (Aguilar et al., 2020; Yu&Tsuei, 2022). Some studies point gamification's effect on peer learning and social interdependence (Yang et al., 2015; Wang et al., 2020). Educational games may also affect students' achievement and emotions, more positively than negatively (Lei et al., 2022). Gamification and new technologies like artificial intelligence (AI) or virtual reality may make teaching theory more exciting (Kriz et al., 2021).

Most games in language education are aimed at facilitating learning foreign languages (Su et al.,2021; Lai&Bower,2019), and only some are explicitly developed for Chinese learning. For instance, Hong et al. (2017) used gamification for recognition of Chinese radicals' structure and found a correlation between intrinsic motivation, online learning self-efficacy, flow experience, and learning progress, whereas Wong&Hsu (2016) noted higher post-test scores and stimulated peer interaction. Tsai et al.(2021) analysed the effectiveness of the Key-Image method - a novel mnemonic (memory aid) tool similar to *Chineasy* method, where characters evolve from a picture in the Chinese learning class. The experimental group outperformed the comparison group and showed greater interest. A similar technique with similar results was used for the *Newby* Chinese game in Australia by Redfern & McCurry (2018) and analysis of games like *Second life* and *Sifteo* cube by Yuan & Wang (2021). Li&Liang (2020) in their study on Chinese secondary school students, state that effectiveness of games in Chinese learning comes with satisfaction from the learning process, which is why gamification can bring students deeper immersion and joy in the language learning process.

Fan, Luo, & Wang (2017) connected Chinese learners with native speakers in the collaborative mobile learning game *ToneWars*, to improve their tone learning. The *Rensselaer Mandarin Project* has been designed in collaboration with IBM for foreign language students to learn Chinese through a virtual visit to China with the use of AI (Allen et al, 2019). Wang, Shi, & Li (2019) discovered the potential of *Wechat* mini games for Chinese learners. Poole et al. (2019) designed the *Mystery Forest* board game for mathematics and Chinese learning for an elementary school in Utah, during which students were eager to use their Chinese language knowledge and communicate with their peers. Chen (2019) developed a Chinese matching game, proving that teachers are capable of designing and using games in the Chinese teaching process. Jamshidifarsani et al.(2019) analysed papers related to technology-based reading intervention programs (also gamification interventions). Wang, Liu, & Zhang (2019) did not find a significant impact on gamification of learning Chinese, but their game Speed Mandarin increased students' confidence in speaking. Chou, Chang, & Hsieh (2020) introduced escape-the-room game with tablets for young Chinese learners. Although the progress of Chinese learning was hard to assess, "motivation was high," and there was peer collaboration. Tang&Taguchi (2021) assigned two groups of Chinese learners from U.S. universities – a Questaurant game group and online lesson groups. Both groups equally improved their results, but the game group had a higher level of motivation. Motivation has increased from games also while studying classical Chinese through e-learning in Lau (2021) and an ancient prose course (Fang&Yang, 2017). He&Loewen (2022) point out that in case of low engagement in second language applications like Memrise teacher support is important. Cho, Andersen, & Kizilcec (2021) developed a game called *Delivery Ghost* for beginner learners of Mandarin, however, the game's interactivity and immersion did not have an impact on learning gains. Wen (2021) use of augmented reality (AR) in the Chinese language learning game developed for schools in Singapore improved the self-learning of students. Fung et al.(2019) use of AR improved Chinese character recognition in Hong Kong, whereas in mainland China, the use of AR for Chinese character recognition brought memorable and joyful results (Wei et al., 2020). Positive impact from the gamification of Chinese language teaching and interventions with mobile Chinese learning games was found in Ying, Yulius, & Juniarto (2021), Ying et al.(2021).

### 3.2. Intelligent Tutoring Systems

Nowadays, emotional ITS can react and adjust to students' motivation and boost their performance. When building the first ITS, a process took around 200 hours of development for each hour of tutored instruction. According to Carnegie Mellon University, the modern ITS may create a 30-minute lesson in about 30 minutes using AI (Spice, 2020). Although the goal of ITS is not to replace the teacher but rather help them with large classes or individualised teaching, some research proves the effectiveness of ITS over human tutoring. ITS has also been used in flipped classroom settings, built into MOOC platforms.

Modern dialogue-driven ITS powered by AI like *Korbit* uses gamification, natural language processing (NLP), machine learning, multimedia for STEM learning (Chen et al., 2020; Serban et al.,2020). When students select a course to enrol and answer a few questions regarding their background, Korbit's outer-loop system decides which exercises to provide for the personalized curriculum. AI-driven system uses data for prediction by using predetermined algorithms. Korbit compares the student's solution with the reference solution using the NLP. If the student provides incorrect information for a question, Korbit's inner-loop system gives some hint. AI helps to receive feedback and understand needs of the learners and select suitable learning methods according to predictive algorithms (Bhutoria, 2022). Browsing through the studies about ITS, one may discover the *Chinese Room Argument*, the philosophical concept of how AI works — searches for answers it does not understand, questions it does not understand, or how to follow instructions. The argument is based on the Searle's example of a native English speaker without Chinese knowledge, who in the room searches for the answers in Chinese to the questions in Chinese, following instructions from people who are outside the room (Kashyap, 2021).

According to Wang (2015), game-based classical Chinese flipped class may positively influence students preparation before class. In his experiment, the learners in the experimental group learned in the flipped classroom with the assistance of ITS. In contrast, the control group was in the flipped classroom and not using ITS. The results showed that all students improved their Chinese knowledge, but learners who used ITS were more motivated in terms of self-directed preview learning, while those using only the traditional textbooks "tended to be more passive." The mobile-assisted learning system also facilitated students'' access to flipped classroom learning.

Chu, Taele, & Hammond (2018) improved the *BopoNoto* sketch recognition technique for Chinese language learners, which is an important technique because a teacher cannot always follow students' writing order of Chinese characters, but the specially designed ITS can. Xu et al.(2021) in their analysis of teaching Chinese characters online, note the interactivity of the ITS. Goksu (2021) names China among the most influential countries in mobile learning, including language learning. Hong et al. (2017) analysed confusion evolving in a game by correcting writing in Chinese. They discovered that confusion is a manageable (by a teacher or ITS) emotion that can be used for error correction in Chinese writing.

## 4. Discussion & Conclusion

Currently, learning foreign languages is undergoing a digital transformation. The aim of the current review was to explore new techniques for learning Chinese by reviewing 2016 to 2022 studies from ScienceDirect and Scopus databases on Chinese learning using games, gamification, or ITS. 7 038 publications from ScienceDirect and Scopus databases were retrieved from which 29 publications were analysed in detail. A few studies are focusing specifically on learning Chinese as a foreign language with the use of games, gamification, or the ITS. Games and ITS described in the current review that might be useful to Chinese learners are: *Chineasy, Chinese-PP, Delivery Ghost, escape-the-room, Key-Image, Memrise, Mystery Forest Newby, Questaurant, Rensselaer Mandarin Project, Second life, Sifteo cube.*

While previous research has established the effectiveness of games, gamification, separately from ITS and without focusing on Chinese language, this study focuses specifically on the review of all three methods in Chinese learning and characteristics (motivation, self-efficacy, progress, effectiveness, learning satisfaction) they are enhancing. The reviews of studies showed that the majority of authors focus on students' motivation, several studies measure it quantitively with different techniques. Many studies address self-efficacy, students 'achievements, learning effectiveness or satisfaction, but a few measure them. Only one study addresses all mentioned characteristics. Hopefully, this study may encourage researchers to reflect upon the different impact technology makes on students, broaden their research questions to several characteristics enhanced by technologies, and not only describe, but also measure the effect of technologies on students.

More in-depth research should explore how games, gamification, and ITS can best be implemented for Chinese language teaching to foreigners.

There are several limitations in this study. First, this review covered studies only from two databases – ScienceDirect and Scopus. Although there are the highest quality journals, but their number is limited. Secondly, the review is limited by a five-year time frame, from 2017 to 2022. Although the analysis is most relevant to new technology, but there could possibly be more analysis made before 2017. Thirdly, reviewed studies are in English. Although some researchers are Chinese, but there is high possibility that a lot of research on Chinese learning is done in Mandarin or other languages. Last but not least, only six characteristics are reviewed in this study, but the impact of technology on students is much broader.

# Annex 1. Literature Review

| Author | Game/ITS | Methodology | Motivation | Self-efficacy | Progress | Effectiveness | Learning satisfaction |
|---|---|---|---|---|---|---|---|
| Tsai et al.(2021) | Chinese radical (key)-image method | Achievement test, inventory | increased | N/A, but efficiency & efficacy increased | | | |
| Li&Liang (2020) | Chinese learning effectiveness | Surveys | | | | 0.807 | 0.802 |
| Fan, Luo, & Wang (2017) | Acquiring Chinese tones through games | Pre- and post-test, survey, interview | gained through confidence | | | proved | present |
| Allen et al.(2019) | Rensselaer Mandarin Project learning & virtual travel (in development) | Descriptive creation of the game | | self-govern | | | |
| Wang, Shi, & Li (2019) | Chinese Language Learning in WeChat Mini programs | Descriptive creation of the game | | | | aimed at, but not measured | |
| Poole et al.(2019) | Collaborative board game | Audio collection & analysis | aimed at, but not measured | aimed at, but not measured | | aimed at, but not measured | aimed at, but not measured |
| Chen (2019) | Chinese matching game | Descriptive creation of the game | aimed at, but not measured | | | | |
| Jamshidifarsani et al.(2019) | Technology-based reading intervention programs | Literature review | aimed at, but not measured | aimed at, but not measured | aimed at, but not measured | aimed at, but not measured | |
| Wang, Liu, & Zhang (2019) | Speed Mandarin computer program | Pre- and post-questionnaire | 3.46-3.5 | aimed at, but not measured | | measured through competences | |
| Chou, Chang, & Hsieh (2020) | Escape-the-room game with tablets | Pre- and post-test, interview | aimed at, but not measured | aimed at, but not measured | | aimed at, but not measured | |
| Tang&Taguchi (2021) | Questaurant game | Recognition & production test, questionnaire | 61.33 for game players vs 52 for no players | | mentioned in the questionnaire | aimed at, but not measured | mentioned in the questionnaire |
| Lau (2021) | E-learning activities in Classic Chinese reading | Pre- and post-questionnaire | 3.19-3.78 | 3.13-3.43 | | aimed at, but not measured | |
| Fang&Yang, 2017 | Avatars and Learning Companions in Studying Chinese Classical Literature | Pre- and post-questionnaire | aimed at, but not measured | | | | |
| He&Loewen (2022) | Memrise | Pre- and post-questionnaire, survey | boosted by 34% | | | supported | |
| Cho, Andersen, & Kizilcec (2021) | Delivery Ghost | Pre- and post-questionnaire, survey | interactivity and immersion are less critical to learning at the beginner-level than a well-structured curriculum | | | | |
| Wen Wen (2018) | Augmented reality enhanced chinese character learning game | Recorded learning process, focus group discussions | engagement | aimed at improved self-learning, but not measured | aimed at, but not measured | | |
| Fung, Fung, & Wan (2019) | Augmented reality and 3D model for children Chinese character recognition | Pre- and post-test, teacher & student focus groups | aimed at, but not measured | aimed at improved self-learning, but not measured | aimed at, but not measured | | |
| Wei et al. (2020) | Mobile AR Laguage Learning Environment Based on Virtual avatar | Pre- and post-test, questionnaire | measured as learning attitude in min, higher with AR (10 min) | | | | higher with AR (4.58 vs 3) |
| Ying, Yulius, & Juniarto (2020) | Chinese learning listening games | Questionnaires | aimed at, but not measured | | aimed at, but not measured | | |
| Ying et al.(2020) | Mandamonic games | Surveys | aimed at, but not measured | | | aimed at, but not measured | |
| Chen et al., 2020 | Games, ITS powered by AI, e.g.Korbit | Literature review | mentioned in 0.91% publications | mentioned in 0.48% publications | | | |
| Serban et al.,2020 | Korbit | Questionnaires | aimed at, but not measured | | average student learning measured as correct answers with pedagogical interventions 39.14% | effective pedagogical interventions | "fun" according to students |
| Bhutoria (2022) | A systematic review of personalized Edtech using AI in the US, China, India | Literature review | aimed at, but not measured | | | aimed at, but not measured | |
| Kashyap, 2021 | Chinese room argument | Literature review | aimed at, but not measured | | aimed at, but not measured | aimed at, but not measured | aimed at, but not measured |

| | | | | | | | | |
|---|---|---|---|---|---|---|---|---|
| Wang, 2015 | Cross-device mobile-assisted classical chinese learning system fo flipped classroom | Questionnaires | enhanced to 4 in comparison to 3.33 | aimed at, but not measured | aimed at, but not measured | aimed at, but not measured | |
| Chu, Taele, & Hammond, 2018 | ITS for correct stroke order in learning Chinese characters | Test, survey | aimed at, but not measured | | aimed at, but not measured | | |
| Xu et al., 2021 | Chinese character online instruction | Questionnaires | aimed at, but not measured | aimed at, but not measured | aimed at, but not measured | aimed at, but not measured | aimed at, but not measured |
| Goksu, 2021 | Bibliometric mapping of mobile learning | Literature review | aimed at, but not measured | | | aimed at, but not measured | |
| Hong et al.,2017 | Game correcting writing in Chinese | Questionnaire | aimed at, but not measured | aimed at, but not measured | | | |